\documentclass[runningheads]{llncs}
\usepackage{cite}
\usepackage{graphicx}
\usepackage{multirow}
\usepackage{amssymb}
\usepackage{array}
\usepackage{url}
\usepackage{enumerate}
\usepackage{enumitem}
\usepackage{amsmath}
\usepackage{booktabs}
\usepackage{multirow}

\usepackage{algorithm}
\usepackage{algpseudocode}
\usepackage{bm}

\usepackage[misc]{ifsym}
\usepackage{lipsum}

\usepackage {xcolor}

\begin{document}

\title{ASC: Appearance and Structure Consistency for Unsupervised Domain Adaptation in Fetal Brain MRI Segmentation}
\titlerunning{ASC: Appearance and Structure Consistency for UDA in Fetal Brain}

%\author{ID 1934}
\author{
Zihang Xu\inst{1}$^{*}$
\and 
Haifan Gong\inst{1,2}$^{*}$
\and 
Xiang Wan\inst{1}
\and 
Haofeng Li\inst{1}$^{(\textrm{\Letter})}$
}
%index{Xu, Zihang}
%index{Gong, Haifan}
%index{Wan, Xiang}
%index{Li, Haofeng}
\authorrunning{Z. Xu et al.}

%
% First names are abbreviated in the running head.
% If there are more than two authors, '\textit{et al.}.' is used.
%
%\institute{Anonymous}
\institute{Shenzhen Research Institute of Big Data, Shenzhen, China 
\and 
The Chinese University of Hong Kong, Shenzhen, China\\
\email{lhaof@sribd.cn}
}

\maketitle              % typeset 

\newcommand\blfootnote[1]{%
\begingroup
\renewcommand\thefootnote{}\footnote{#1}%
\addtocounter{footnote}{-4}%
\endgroup
}

\begin{abstract}
Automatic tissue segmentation of fetal brain images is essential for the quantitative analysis of prenatal neurodevelopment. However, producing voxel-level annotations of fetal brain imaging is time-consuming and expensive. To reduce labeling costs, we propose a practical unsupervised domain adaptation (UDA) setting that adapts the segmentation labels of high-quality fetal brain atlases to unlabeled fetal brain MRI data from another domain. To address the task, we propose a new UDA framework based on Appearance and Structure Consistency, named ASC. We adapt the segmentation model to the appearances of different domains by constraining the consistency before and after a frequency-based image transformation, which is to swap the appearance between brain MRI data and atlases. Consider that even in the same domain, the fetal brain images of different gestational ages could have significant variations in the anatomical structures. To make the model adapt to the structural variations in the target domain, we further encourage prediction consistency under different structural perturbations. Extensive experiments on FeTA 2021 benchmark demonstrate the effectiveness of our ASC in comparison to registration-based, semi-supervised learning-based, and existing UDA-based methods. 
\blfootnote{This work is supported by Chinese Key-Area Research and Development Program of Guangdong Province (2020B0101350001), 
and the National Natural Science Foundation of China (No.62102267), 
and the Guangdong Basic and Applied Basic Research Foundation (2023A1515011464), 
and the Shenzhen Science and Technology Program (JCYJ20220818103001002), 
and the Guangdong Provincial Key Laboratory of Big Data Computing, The Chinese University of Hong Kong, Shenzhen. 

Haofeng Li is the corresponding author. %(lhaof@sribd.cn). 
Zihang Xu and Haifan Gong contribute equally to this work.
}
\end{abstract}

\section{Introduction}
Magnetic resonance imaging (MRI) has emerged as an important tool for assessing brain development in utero \cite{de2022adverse,hart2020accuracy,benkarim2017toward,li2022view}. Since manual segmentation is time-consuming~\cite{zhou2021ssmd} and suffers from high inter-rater variability in quantitative assessment, automatically segmenting brain tissue from MRI data becomes an urgent need \cite{gousias2012magnetic,makropoulos2018review,huang2022attentive}. However, the available annotated fetal brain datasets are limited in number and heterogeneity, hindering the development of automatic strategy. 

To achieve the unsupervised fetal brain tissue segmentation, registration-based methods~\cite{wang2012multi,sabuncu2010generative} use image registration and label fusion to obtain the segmentation result from a set of templates \cite{sanroma2018learning,xie2023deep}. Still, the accuracy of these methods is not sufficient due to the complexity of registration, and they usually underperform on the abnormal fetal brain image. 
\begin{figure*}[!tbp]
\includegraphics[width=\textwidth]{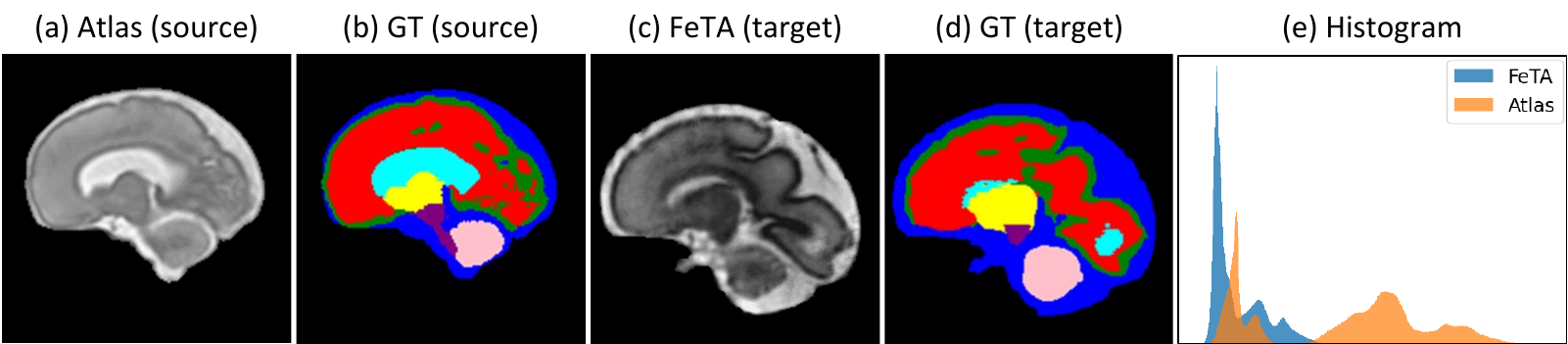}
\centering
\caption{Comparison of MR images from brain atlases (a) and FeTA dataset (c). (b) and (d) show the ground truth. (e) shows the gray-scale histograms of the samples of 30-week gestational age from the FeTA dataset and brain atlases. Horizontal and vertical coordinates denote intensity values and voxel numbers, respectively. 
}
\label{fig0}
\end{figure*}
Recently, deep learning (DL) based unsupervised domain adaptation (UDA) methods have shown their advance on medical image segmentation tasks \cite{chen2020unsupervised,al2021olva,han2021deep}. UDA methods usually narrow the distribution discrepancy between source and target domains by enforcing image-/feature-level alignment \cite{xie2022unsupervised,tomar2021self,huo2018synseg,yang2020fda}. 
In addition to inter-domain knowledge transfer, some works \cite{li2020dual,zhao2021mt} explored the knowledge from both intermediate domains. 

However, existing UDA methods in medical imaging mainly focus on the gap from different modalities (e.g., CT and MR), and pay less attention to the domain gap from different centres. 
Due to motion artifacts~\cite{payette2021automatic}, it is difficult to collect high-quality fetal brain MR images and is expensive to label the newly collected data voxel-wisely. The above observations motivate us to establish a new UDA problem setting that aims to transfer the segmentation knowledge from the publicly available atlases to unlabeled fetal brain MRIs from new centres. 

To solve the above UDA task, we propose an Appearance and Structure Consistency (ASC) framework. Consider the fact \cite{yang2020fda,yue2021robust} that swapping the low-level spectrum between images can exchange their style/color/brightness without changing semantic content,  while swapping the higher spectrum introduces unwanted artifacts. Thus, we propose to align appearance by only swapping the low-level spectrum. 
We develop an appearance consistency regularization based on a frequency-based appearance transformation, which is performed between labeled source data and unlabeled target data. Specifically, the source domain and the source data under the target appearance, are supervised with the same source labels. Then, the target data under the target appearance and source appearance are forced to maintain the same segmentation via dual unsupervised appearance consistency. Considering that significant variances in the shape of abnormal fetal brain tissue can cause difficulties in segmentation, we further constrain structure consistency under different perturbations in the target domain, besides aligning the inter-domain appearance gap. All the above consistency constraints are integrated with a teacher-student framework.

The contributions of this work are three-fold: (1) we propose a novel Appearance and Structure Consistency framework for UDA in fetal brain tissue segmentation; 
(2) we propose to address a practical UDA task adapting publicly available brain atlases to unlabeled fetal brain MR images; 
(3) experimental results on FeTA2021 benchmark \cite{payette2021automatic} show that the proposed framework outperforms representative state-of-the-art methods.

\begin{figure*}[tbp]
\includegraphics[width=\textwidth]{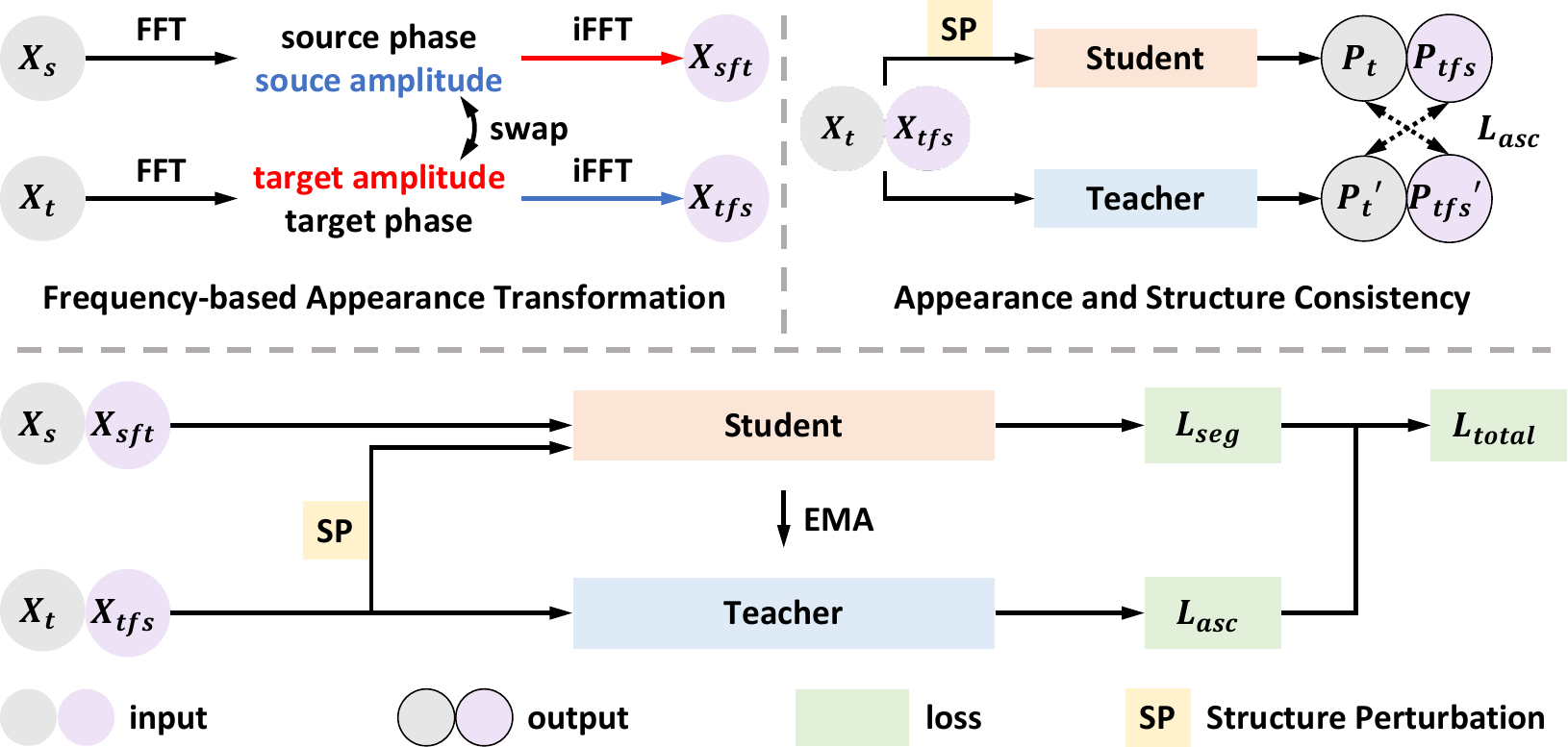}
\caption{Overview of the proposed Appearance and Structure Consistency framework. The student model learns from source data $X_{s}$ and frequency-based transformed source data $X_{sft}$ via the supervised loss $L_{seg}$. The appearance and structure consistency is achieved by the loss $L_{asc}$.} 
\label{fig1}
\end{figure*}

\section{Methodology}
In the UDA setting, $D_{s}=\{X_{s}, Y_{s}\}^{M}_{s=1}$ denotes a set of source domain images (e.g., fetal brain atlases) and corresponding labels, respectively. $D_{t}=\{X_{t}\}^N_{t=1}$ denotes a set of target domain images (e.g., images from the FeTA benchmark). We aim to learn a semantic segmentation model for target domain data based on the labeled source and unlabeled target domain data. Usually, this goal is achieved by minimizing the domain gap between source domain samples $D_{s}$ and target domain samples $D_{t}$. Fig.\ref{fig1} depicts the proposed Appearance and Structure Consistency (ASC) framework based on a teacher-student model. 

\subsection{Frequency-based Appearance Transformation}
\begin{figure}[tbp]
\includegraphics[width=0.9\textwidth]{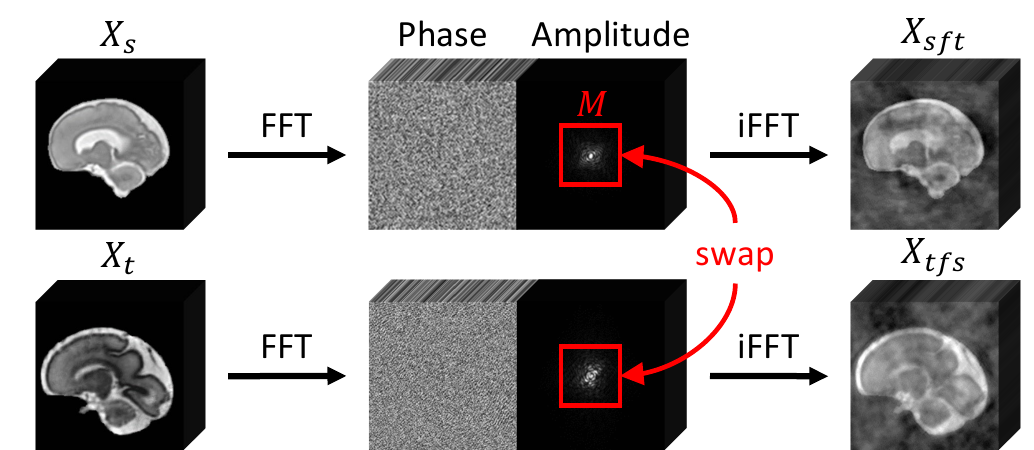}
\centering
\caption{Illustration of the frequency-based appearance transformation. The transformation exchanges the domain-specific appearance between two images by swapping their low-frequency components of the spectrum. 
FFT is the Fast Fourier Transform. Phase component and amplitude spectrum are denoted by $\mathcal{F}^{P}$ and $\mathcal{F}^{A}$ in Eq.~(\ref{eq2}), respectively.}
\label{fig2}
\end{figure}

Atlases are magnetic resonance fetal images with ``average shape''. Domain shifts between the atlases and fetal images are mainly due to the texture, different hospital sensors, illumination or other low-level sources of variability. However, traditional UDA employing GAN to synthetic style-transfer images hardly capture such domain shift. Thus, we align the low-level statistics based on Fourier transformation to narrow the distribution of the two domains. This process is shown in Fig.\ref{fig2}. Taking source data as an example, we compute the Fast Fourier transform (FFT) of each input image to obtain an amplitude spectrum $\mathcal{F}^{A}$ and a phase component $\mathcal{F}^{P}$, where the low-frequency part of the amplitude of the source image $\mathcal{F}^{A}(X_{s})$ is swapped with the amplitude of the target image $\mathcal{F}^{A}(X_{t})$. Then, the transformed spectral representation of $X_{s}$ and the original phase $\mathcal{F}^{P}(X_{s})$ are mapped back to the image $X_{sft}$ by inverse FFT (iFFT). $X_{sft}$ has the same content as $X_{s}$ and similar appearance to $X_{t}$. The above process can be formally defined as:
\begin{equation}
\mathcal{F}^{A}(X_{sft})=M\cdot \mathcal{F}^{A}(X_{t})+(1-M)\cdot \mathcal{F}^{A}(X_{s}),
\end{equation}
\begin{equation}
X_{sft}=\mathcal{F}^{-1}([\mathcal{F}^{A}(X_{sft}),\mathcal{F}^{P}(X_{s})]),
\label{eq2}
\end{equation}
where the mask $M=\mathcal{I}_{(h,w,d)\in[-\beta H:\beta H,-\beta W:\beta W,-\beta D:\beta D]}$ controls the proportion of the swapped part over the whole amplitude by a parameter $\beta \in (0,1)$. Here we assume the center of the image is (0, 0, 0). Then we can train a student network with domain alignment images $X_{sft}$, the original images $X_{s}$ and the labels $Y_{s}$ by minimizing the dice loss:
\begin{equation}
L_{seg} = L_{dice}(P_{s},Y_{s})+L_{dice}(P_{sft},Y_{s}),
\end{equation}
where $P_{s}$ and $P_{sft}$ are the prediction of $X_{s}$ and $X_{sft}$, respectively. 

\subsection{Appearance Consistency}
The above loss function imposes an implicit regularization before and after frequency-based transformation. In other words, source domain image $X_{s}$ and its transformation image $X_{sft}$ should predict the same segmentation. However, the label of images from the target domain is not available $X_{t}$ in UDA settings. As a replacement, we propose a teacher model for keeping semantic consistency across domain transformation. Specifically, the target domain image $X_{t}$ and its aligned image $X_{tfs}$ are regarded as representations of an object under different domains. Given the inputs of $X_{t}$ and $X_{tfs}$ of teacher and student models, we expect their predictions to be consistent. 
Further, considering that appearance transformation may break certain semantic information and make the model learn the wrong mapping relationship, we employ a form of dual consistency, which directs the model to focus on invariant information between the two views. $f(\cdot)$ and $f'(\cdot)$ represent the outputs of the student model and the teacher model, respectively. Following the conventional consistency learning methods \cite{tarvainen2017mean}, we calculate the appearance consistency loss $L^{app}_{con}$ between the teacher and student networks as:
\begin{equation}
L^{app}_{con}=\frac{1}{N}\sum^{N}_{i=1}||f(X_{t,i})-f'(X_{tfs,i})||^2+\frac{1}{N}\sum^{N}_{i=1}||f(X_{tfs,i})-f'(X_{t,i})||^2.
\end{equation}

\subsection{Structure Consistency}
Although frequency-based transformation and appearance consistency align the two domains' styles, the variance of tissue structure in pathological subjects still brings difficulty to domain alignment, which limits the model's gener alization ability. To this end, we utilise the teacher-student model \cite{tarvainen2017mean} keeping prediction consistency $L^{str}_{con}$ under structure perturbation \cite{yun2019cutmix} to alleviate the above problem. Here structure perturbation is $sp$ for short. To achieve the structure perturbation, we first use a 3D cuboid mask consisting of a single box that randomly covers 25-50\% of the image area at a random position, to blend two input images, which are sampled from the same batch.  Then we blend the teacher predictions for the input images to produce a pseudo label for the student prediction of the blended image.  Such an operation changes the original structural information, reduces the overfitting risk, and increases the robustness of the model to adapt to different structural variations. As appearance transformation doesn't affect the structure information, we add $sp$ to both $X_{t}$ and $X_{tfs}$ to obtain  $X_{t,sp}$ and $X_{tfs,sp}$, which are fed into the teacher-student model and expected their predictions to be consistent. Then, $L^{str}_{con}$ and $L^{app}_{con}$ are combined as $L_{asc}$: 
\begin{equation}
L_{asc}=\frac{1}{N}\sum^{N}_{i=1}||f(X_{t,i,sp})-f'(X_{tfs,i})||^2+\frac{1}{N}\sum^{N}_{i=1}||f(X_{tfs,i,sp})-f'(X_{t,i})||^2.
\end{equation}
\subsection{Overall Training Strategy}
We calculate appearance and structure consistency using the same teacher model. Its model weight $\theta'$ is updated with the exponential moving average (EMA) of the student model $f(\theta)$, i.e., $\theta'_{t}= \alpha\theta'_{t-1} + (1-\alpha) \theta_{t}$, where $\alpha$ is the EMA decay rate that reflects the influence level of the current student model parameters.

Let $\lambda$ control the trade-off between the supervised loss and the unsupervised regularization loss, the overall loss is: 
\begin{equation}
L_{total} = L_{seg} + \lambda L_{asc}.
\end{equation}

\section{Experiment}
\begin{table}[tbp]
\centering
\caption{Comparison with the state-of-the-art methods. DSC$_{A}$ / DSC$_{N}$ stands for the average DSC calculated in the Abnormal / Normal subset. The best results are in \textbf{bold}. The DSC of each category is in the supplementary material.}
\begin{tabular}{@{}cccccc@{}}
\toprule
Strategy & Method & Venue & DSC$_{A}$ & DSC$_{N}$ & Avg. DSC\\ 
\hline
Upper  & Supervised ($D_{t}$)&  -  & $81.4_{\pm0.2}$   &$83.1_{\pm0.3}$   &$82.0_{\pm0.1}$  \\
\hline
Lower  & W/o Adaptation&  -  & $73.1_{\pm0.6}$  & $79.0_{\pm0.3}$  & $75.3_{\pm0.5}$\\ 
\hline
Registration-based& SCALE \cite{sanroma2018learning}&  MIA'18   & $63.6_{\pm1.5}$    & $77.3_{\pm1.7}$    & $68.7_{\pm1.7}$\\
\hline
\multirow{3}{*}{UDA}     & FDA \cite{yang2020fda}&  CVPR'20     & $74.4_{\pm0.4}$    & $80.4_{\pm0.4}$    & $76.7_{\pm0.4}$\\
& OLVA \cite{al2021olva}&  MICCAI'21  & $73.3_{\pm0.6}$    & $79.1_{\pm0.3}$    & $75.6_{\pm0.3}$ \\
& DSA \cite{han2021deep}&  TMI'22  & $73.4_{\pm0.6}$    & $79.9_{\pm0.4}$    & $75.9_{\pm0.5}$ \\
\hline
\multirow{2}{*}{SSL}     & CUTMIX \cite{yun2019cutmix}&  ICCV'19  & $74.1_{\pm0.1}$    & $79.7_{\pm0.5}$    & $76.2_{\pm0.2}$ \\
& ASE-NET \cite{lei2022semi}&  TMI'22  & $73.7_{\pm0.3}$    & $79.8_{\pm0.3}$    & $76.2_{\pm0.2}$\\ 
\hline
UDA& ASC (ours)&  -   & $\bm{76.6_{\pm0.2}}$    & $\bm{81.7_{\pm0.1}}$    & $\bm{78.5_{\pm0.1}}$\\
\bottomrule
\end{tabular}

\label{exp:sota}
\end{table}

\subsection{Dataset and pre-processing}
We evaluated our method on the Fetal Brain Tissue Annotation and Segmentation Challenge (FeTA) 2021 benchmark dataset \cite{payette2021automatic}, which contains 80 3D T2 MRI volumes with manual segmentations annotation of external cerebrospinal fluid (eCSF), grey matter (GM), white matter (WM), lateral ventricles (LV), cerebellum (CBM), deep grey matter (dGM) and brainstem (BS). The dataset cohort consisted of two subgroups: 31 neurological fetuses and 49 fetuses with abnormal development. Following the general UDA setting \cite{huo2018synseg}, the target set was randomly divided into 40 scans for training and 40 scans for testing. A collection from three atlases was used as the source set, including 32 neurotypical fetal brain atlases \cite{gholipour2017normative,wu2021age} and 15 spina bifida fetal brain atlases \cite{fidon2022spatio}. Segmentations for all tissue types are available for all the atlas data. Some examples of slices, segmentations and histogram distribution are shown in Fig.~\ref{fig0}. 
We cropped the foreground region of fetal volumes and reshaped them to $144\times144\times144$. Before being fed into the network, the input scans were normalized to a zero mean and unit variance.

\subsection{Implementation details}
All models were implemented in PyTorch 1.12 and trained with NVIDIA A100 GPU with CUDA 11.3. Following the top-ranked method in the FeTA2021 competition \cite{payette2021automatic}, we use SegResNet \cite{myronenko20193d} as the backbone for the teacher/student model. 
The network parameters were optimized with Adam with the initial learning rate of $1 \times 10^{-4}$. ``Poly'' learning rate policy is applied, where $lr=lr_{init}\times(1-\frac{epoch}{epoch_{total}})^{{0.9}}$. The batch size and training epoch were set to 4 (2 from each domain) and 100, respectively. The EMA decay rate $\alpha$ of the teacher model was set to 0.99, and hyperparameters $\lambda$ were ramped up individually with function $\lambda(t) = \gamma \times e^{(-5(1-\frac{t}{t_{max}})^{2})} $, where $t$, $t_{max}$ and $\gamma$ were the current step, the last step and weight, respectively. $\beta$ was set to 0.1. Cutmix\cite{yun2019cutmix} was used for target images as the structure perturbation for consistency regularization. We employed the student model prediction as the final result and used the Dice Similarity Coefficient (DSC) scores to evaluate the accuracy of the results. The average results of three runs were reported in all experiments.

\subsection{Comparison with the State-of-the-arts}
\begin{figure}[tbp]
\includegraphics[width=\textwidth]{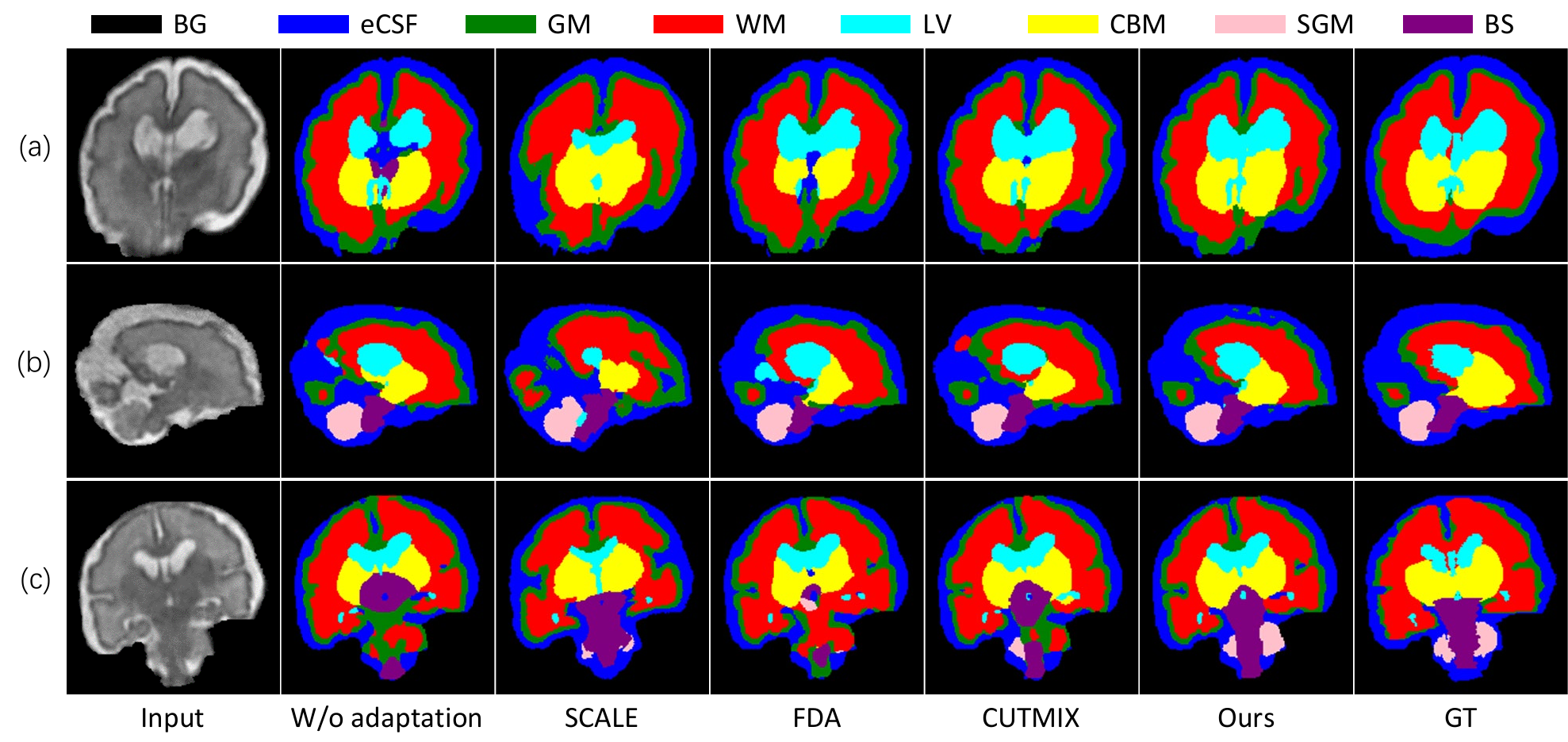}
\caption{Visual comparisons with existing methods. Due to space limitations, we only present those with the best average DSC in Registration-based, UDA and SSL methods. It can be seen that our predictions are visually closer to the ground truth than the other methods.}
\label{fig3}
\end{figure}
We implemented several state-of-the-art label-limited segmentation methods for comparison, including Registration-based (SCALE\cite{sanroma2018learning}), Unsupervised Domain Adaptatopm (UDA) (FDA \cite{yang2020fda}, OLVA\cite{al2021olva} and DSA\cite{han2021deep}) and Semi-supervised Learning (CUTMIX\cite{yun2019cutmix}, ASE-NET\cite{lei2022semi}) in Table \ref{exp:sota}. It reports the segmentation performance of UDA of adapting atlas to the fetal brain on FeTA2021, including average DSC for the full set, normal set and abnormal set. The upper bound is given by supervised training which uses fully-labeled target data for model training. 

It is worth noting that registration-based \cite{sanroma2018learning} only successfully segments on the normal set, and GAN-based UDA approach \cite{han2021deep} performs worse than the frequency-based approach \cite{yang2020fda} on the abnormal set. The proposed method is superior to the existing state-of-the-art methods \cite{yang2020fda,yun2019cutmix} and achieves mean Dice of 78.5\% over the seven tissue structures, reducing the Dice gap to supervised training to 3.5\%. Compared with the baseline model (W/o adaptation), our proposed learning strategy further improves the performance by an average of 3.2\% Dice. Visual results in Fig.\ref{fig3} show our method can perform better in the junction areas of brain tissue.

\subsection{Ablation Study and Sensitivity Analysis}
The ablation study is shown in Table \ref{abla}, and all the results boost our method's performance. ``M1'' represents the lower bound that only trains on the source domain data $D_{s}$. ``M2'' uses the aligned source images $X_{sft}$ for training. The following component are based on $L_{asc}$, which is decoupled as $L^{app}_{con(X_{t})}$, $L^{app}_{con(X_{tfs})}$ and $L^{str}_{con}$. ``M3'' denotes the appearance consistency loss $L^{app}_{con(X_{t})}$ to align distribution from source to target. ``M4'' indicates the dual-view appearance consistency loss to constrain semantic invariance. ``M5'' denotes the structure consistency $L^{str}_{con}$. 

We can see that appearance consistency can boost performance on normal and abnormal fetal MRIs, showing that minimizing the appearance gap between the source domain and target domain is effective. Further, we can obtain better results on the abnormal samples by applying the structure consistency loss. 
Table \ref{hyp} shows that the performance of our method grows with the increase of the hyper-parameter $\gamma$ of consistency loss, and achieves the best when $\gamma=200$. Besides, \textbf{efficiency analysis} is shown in the supplementary material.

\begin{table}[tbp]
\centering
\caption{Ablation study of the contribution of each component in the proposed framework. The highest evaluation score is marked in \textbf{bold}.}
\begin{tabular}{@{}ccccccccc@{}}
\toprule 
\multirow{2}{*}{Method}   &  \multirow{2}{*}{$L_{seg(X_{s})}$}   &  \multirow{2}{*}{$L_{seg(X_{sft})}$}  & \multicolumn{3}{c}{$L_{asc}$}  &  \multirow{2}{*}{DSC$_{A}$}  &  \multirow{2}{*}{DSC$_{N}$}  &  
\multirow{2}{*}{Avg DSC}   \\ 
\cline{4-6}
&&&  $L^{app}_{con(X_{t})}$  &  $L^{app}_{con(X_{tfs})}$  &  $L^{str}_{con}$  &    \\
\hline
M1       &$\surd$&       &       &       &       &$73.1_{\pm0.6}$&$79.0_{\pm0.3}$&$75.3_{\pm0.5}$\\ 
M2       &$\surd$&$\surd$&       &       &       &$74.3_{\pm0.2}$&$80.4_{\pm0.3}$&$76.6_{\pm0.2}$\\ 
M3       &$\surd$&$\surd$&$\surd$&       &       &$75.6_{\pm0.5}$&$80.9_{\pm0.2}$&$77.3_{\pm0.3}$\\ 
M4       &$\surd$&$\surd$&$\surd$&$\surd$&       &$75.8_{\pm0.1}$&$81.4_{\pm0.3}$&$77.9_{\pm0.2}$\\ 
M5       &$\surd$&$\surd$&$\surd$&$\surd$&$\surd$&$\bm{76.6_{\pm0.2}}$&$\bm{81.7_{\pm0.1}}$&$\bm{78.5_{\pm0.1}}$\\
\bottomrule 
\end{tabular}
\label{abla}
\end{table}

\begin{table}[!tbp]
\centering
\caption{Performance of our method with different values of hyperparameters $\gamma$, which are used to balance consistency loss and supervisory loss.}
\setlength{\tabcolsep}{3mm}{
\begin{tabular}{@{}l|ccccc@{}}
\toprule 
$\gamma$ value  &       10       &  100           &  200           &  500          &  1000  \\ 
\hline
Avg. DSC         &$78.1_{\pm0.1}$&$78.4_{\pm0.1}$&$\bm{78.5_{\pm0.1}}$&$78.4_{\pm0.2}$&$78.2_{\pm0.4}$\\
\bottomrule
\end{tabular}
}
\label{hyp}
\end{table}

\section{Conclusion}
In this paper, we present a novel UDA framework and an atlas-based UDA setting for fetal brain tissue segmentation. Our method integrates appearance consistency encouraging the model to adapt different domain styles to narrow the domain gap and structure consistency making the model robust against the anatomical variations in the target domain. 
Experiments on the FeTA2021 benchmark demonstrate that our method outperforms the state-of-the-art methods. The proposed novel setting of atlas-based UDA could provide accurate segmentation for the fetal brain MRI data without pixel-wise annotations, greatly reducing the labeling costs.

\bibliographystyle{splncs04}
\bibliography{ref}
\end{document}